\documentclass[submission]{eptcs}
\usepackage{breakurl}             

\usepackage{tabularx,colortbl}
\usepackage[table]{xcolor}
\usepackage[pdftex]{graphicx}
\usepackage{subfigure}
\usepackage{amsmath}
\usepackage{amssymb}
\usepackage{amsthm}
\usepackage{mathrsfs}
\usepackage{appendix}
\usepackage{algorithm}
\usepackage{algpseudocode}
\usepackage{xypic}
\usepackage{subfigure}
\usepackage{multirow}
\usepackage{wrapfig}
\usepackage{enumerate}

\usepackage{ifthen} 
\usepackage{xifthen} 

\newtheorem{definition}{Definition}
\newtheorem{theorem}{Theorem}
\newtheorem{lemma}{Lemma}

\title{A Unifying Approach to Decide Relations for Timed Automata and their Game Characterization}
\author{
\hspace{.4in}Shibashis Guha \footnote{The research of Shibashis Guha was supported by Microsoft Corporation and Microsoft Research India under the Microsoft Research India PhD Fellowship Award.} 
\quad \quad \quad \quad \quad \qquad \hspace{.2in} Shankara Narayanan Krishna
\institute{Indian Institute of Technology Delhi \quad \quad \qquad \qquad \qquad Indian Institute of Technology Bombay}
\and
Chinmay Narayan \quad \qquad \qquad \qquad \qquad \qquad S. Arun-Kumar
\institute{Indian Institute of Technology Delhi \quad \quad \qquad \qquad \qquad Indian Institute of Technology Delhi}
}
\def\titlerunning{Relations for Timed Automata}
\def\authorrunning{S. Guha, S. N. Krishna, C. Narayan \& S. Arun-Kumar}

\begin{document}
\newcommand{\track}[1]{{\textcolor{red}{#1}}}
\newcommand{\remove}[1]{{\textcolor{green}{#1}}}
\newcommand{\delete}[1]{{}}
\newcommand{\sembrack}[1]{[\![#1]\!]}
\newcommand{\existsdelay}{\exists \!\!\!\! \exists}
\newcommand{\foralldelay}{\forall \!\!\!\! \forall}
\newcommand{\realpos}{\mathbb{R}_{\ge 0}}
\renewcommand{\to}[1]{\xrightarrow{#1}}
\newcommand{\true} {\mbox{\texttt{t\!t}}}
\newcommand{\false} {\mbox{\texttt{f\!f}}}
\newcommand{\isin} {\:\mbox{\texttt{\underline{in}}}\:}
\newcommand{\mlineqn}[5]{\begin{center}\ensuremath{#1 = \left\{ 
  \begin{array}{l l}
    #2 & \quad \text{#3}\\
    #4 & \quad \text{#5}
  \end{array} \right.}\end{center}}

\newcommand{\TGame}[3]{%
\ifthenelse{\isempty{#1}}
{\ifthenelse{\isempty{#3}}{\Gamma_{#2}^{{#3}}}{\Gamma_{#2}^{{#3}}}}               
            {#1\!-\!\Gamma_{#2}^{{#3}}}    
}

\newenvironment{keywords}{
       \list{}{\advance\topsep by0.35cm\relax\small
       \leftmargin=1cm
       \labelwidth=0.35cm
       \listparindent=0.35cm
       \itemindent\listparindent
       \rightmargin\leftmargin}\item[\hskip\labelsep
                                     \bfseries Keywords:]}
     {\endlist}
\newtheorem{fact}{Fact}
\newtheorem{corollary}{Corollary}

\maketitle

\begin{abstract}
In this paper we present a unifying approach for deciding various bisimulations, simulation equivalences and preorders between two timed automata states. We propose a zone based method for deciding these relations in which we eliminate an explicit product construction of the region graphs or the zone graphs as in the classical methods. Our method is also generic and can be used to decide several timed relations. We also present a game characterization for these timed relations and show that the game hierarchy reflects the hierarchy of the timed relations. One can obtain an infinite game hierarchy and thus the game characterization further indicates the possibility of defining new timed relations which have not been studied yet. The game characterization also helps us to come up with a formula which encodes the separation between two states that are not timed bisimilar. Such distinguishing formulae can also be generated for many relations other than timed bisimilarity.
\end{abstract}
\section{Introduction} \label{sec-intro}
Bisimulation \cite{RM1} is one of the most important 
notions used to study process equivalence in concurrency theory. 
Given two processes (untimed/timed/probabilistic), deciding whether 
they are equivalent in some way\delete{or the other} is a fundamental question of practical significance; 
over the years, several researchers have contributed theory and techniques to answer this question. 
In this paper, we are interested in checking various kinds of equivalences and preorders between timed systems. 

Timed automata, introduced in \cite{AD1} are one of the most popular formalisms for modelling timed systems. 
It is known that given two timed automata\delete{$A$ and $B$},
checking whether they accept the same timed language\delete{$L(A)=L(B)$} is undecidable \cite{AD1}. 
However, bisimulation equivalences between timed automata have been shown to be
decidable \cite{LAKJ1}\cite{KC1}\cite{LY1}\cite{WL1}. 
The decidability of timed bisimilarity between two timed automata 
was proved in \cite{KC1} via a product construction on region graphs. 
 \cite{LLW1} also uses regions as the basis of checking timed bisimilarity for timed automata.
To overcome the state space explosion in region graphs, \cite{WL1} applies the product 
construction on zone graphs. The article \cite{TY1} proposes weaker equivalences (several 
variants of time abstracted bisimulations), and 
uses zone graphs for the same purpose of overcoming the state explosion in region graphs. 

In this work, we propose a uniform way of deciding various timed and time abstracted relations
present in the literature using a zone based approach. The zone graph is constructed in such a way that
every zone is (i) convex, and (ii) intersects with exactly one hyperplane on elapsing time.
First, for deciding timed bisimilarity, we define \emph{corner point bisimulation} and prove that 
two timed automata states are corner point bisimilar iff they are timed bisimilar.  
Apart from the fact that ours is a zone based approach, we also 
do not compute a product of individual zone graphs, as done in \cite{WL1}.
Thus we expect our approach to save computation since it does not require the product zone 
graph to be stored along with the individual zone graphs of the two timed automata.\delete{First,  we define \emph{corner point bisimulation} and prove that 
two timed automata are corner point bisimilar iff they are timed bisimilar.  
For checking the corner point equivalence, we construct a zone graph in which 
every zone (i) is convex, and (ii) intersects with exactly one hyperplane on elapsing time. 
 Apart from the fact that ours is a zone based approach, we also 
do not compute a product of individual zone graphs, as done in \cite{WL1}.
This makes the computation more efficient compared to existing approaches where the product 
of individual zone graphs, along with the component zone graphs need to be stored,  for checking timed bisimulation.
}
Moreover, the product based approach cannot be used to check all possible relations, for instance, 
it is not useful in checking timed performance prebisimulation \cite{GNA1}. Corresponding to each of the 
bisimulation relations described above, we can consider a simulation relation and
our zone graph can be used to check all\delete{of} these relations in a uniform way, 
Further, our method checks timed bisimulation between two states with arbitrary rational valuations; many of the existing approaches 
\cite{LLW1}, can only  check for timed bisimulation between the initial states.

Next, we define a game semantics corresponding to the various timed relations; 
this is an extension of Stirling's bisimulation games for discrete time relations \cite{CS1}. 
The game theoretic formulation obviates the need for tedious operational reasoning
which is required many a time to compare various timed relations: 
the game formulation helps in obtaining a hierarchy among various timed relations in a very elegant and succinct way. 
Playing these games on two timed automata which are not timed bisimilar, 
we synthesize a formula which captures the difference.
The technique of synthesizing distinguishing formulae on two structures using EF games 
is known in the literature \cite{HS1}. 
Given two timed automata $A$ and $B$,  \cite{LLW1} builds a characteristic formula 
$\psi_A$ that describes $A$ and checks if $B \models \psi_A$; $A$ and $B$ are 
timed bisimilar iff $B \models \psi_A$. The distinguishing formula $\varphi$ we synthesize, only captures 
the difference between $A$ and $B$; for many practical situations, $\varphi$ 
would hence be much more succinct than $\psi_A$.
Paper \cite{GL1} also describes a method for constructing a distinguishing formula. However, there too the formula construction depends on the entire (branching) structure of a timed automaton, whereas in our method, the formula is synthesized based on the moves in the game and thus leads to a more succinct formula.
Given a specification $S$, and an implementation $I$, both modeled using timed automata,
 our approach can be used to synthesize the distinguishing formula $\varphi$ (if it exists); 
 $\varphi$  can then be used to refine $I$ to obtain an implementation $J$ which satisfies $S$.   
A prototype tool which constructs the zone graph as described above, and checks for various timed relations 
is underway.\delete{Our tool also implements the game based approach, and we look forward to use it to generate a distinguishing formula that will\delete{uses the distinguishing formula
to} refine implementations that do not agree with the specification.}
Our tool thus will be a unifying framework to check various timed and time abstracted relations; it
will also aid in system refinement by generating a distinguishing formula.

In section \ref{sec-ta}, we give a brief
introduction to timed automata, introduce several definitions required in the paper and describe 
the way we construct the zone graph\delete{, which is termed a \emph{zone valuation graph} \cite{GNA1}}. 
In section \ref{sec-timedrels} we describe the various timed and time abstracted relations considered in 
this work. In section \ref{sec-dec}, we present the methods for deciding these\delete{various timed} relations.
The game semantics is given in section \ref{sec-game}. The zone \delete{valuation graph can be}graph construction used here acts as a common framework to decide several kinds of timed and time abstracted  relations. Finally, we conclude in section \ref{sec-conc}.

\section{Timed Automata}\label{sec-ta}
\emph{Timed automata}, introduced in \cite{AD1} are a very popular 
formalism for modelling time critical systems. 
These are finite state automata over which time constraints 
are specified using real variables called {\it clocks}.
Given a finite set of clocks $C$, the set of constraints $\mathcal{B}(C)$ allowed are 
given by the grammar $g ::= \; x \smile c \:|\: g \wedge g$,
where $c \in \mathbb{N}$ and $x \in C$ and $\smile \: \in \: \{\le, <, =, >, \ge\}$.
Formally a \emph{timed automaton} is a tuple $A=(L, Act, l_0, E, C)$ where
(i) $L$ is a finite set of locations, (ii) $Act$ is a finite set 
of visible actions, (iii) $l_0 \in L$ is the initial location, 
and (iv) 
$E \subseteq L \:\times\: \mathcal{B}(C) \:\times\: Act \:\times\: 2^C \:\times\: L$ is a finite set of \emph{edges}.
Given two locations $l, l'$, a transition from $l$ to $l'$ is of the form 
$(l, g, a, R, l')$: on action $a$, we can go from $l$ to $l'$ if the constraints 
specified by $g$ are satisfied; $R \subseteq C$ is a set of clocks which are reset to zero during  
the transition.

\subsection{Semantics} The semantics of a timed automaton can be described with a \emph{timed labeled
transition system} (TLTS) \cite{LAKJ1}. Let $A= \: (L, Act, l_0, E, C)$ be a timed automaton over a set of clocks
$C$ and a set of visible actions $Act$. The timed transition system $T(A)$ generated by $A$ can be defined as 
$T(A) = (Q, Lab, Q_0, \{\stackrel{\alpha}{\longrightarrow} | \alpha \in Lab\})$, where 
$Q \: = \: \{(l,v)\:|\: l \in L, v \in {\realpos}^{|C|}\}$ is the set of  \emph{states}; each state is of the form $(l,v)$, where $l$ is a location of the timed automaton 
and $v$ is a valuation assigned to 
the clocks of $A$. $Lab = Act \cup \realpos$ is the set of labels.  
Let $v_0$ denote the valuation such that $v_0(x)=0$ for all 
$x \in C$. $Q_0=(l_0, v_0)$ is the initial state of $T(A)$.
A transition happens in one of the following ways:\\ 
(i) Delay transitions : $(l,v)\stackrel{d}{\longrightarrow}(l, v+d)$. Here, $d \in \realpos$ and $v + d$ is the valuation
in which the value of every clock\delete{value} is incremented by $d$. \\
(ii) Discrete transitions : $(l,v) \stackrel{a}{\longrightarrow} (l', v')$ if for an edge 
$e=(l,g,a,R,l') \in \: E$, $v \models g, v' = v_{[R \leftarrow 0]}$, where $v_{[R \leftarrow 0]}$ denotes that the valuation of every clock in $R$ has been reset to 0, while the remaining clocks are unchanged.
From a state $(l,v)$, we take an $a$-transition \delete{$e$}to reach a state $(l', v')$
if the valuation $v$ of the clocks satisfies $g$; 
after this,  the clocks in $R$ are reset while those in $C\backslash R$ remain unchanged. 

For example, let $A$ be a timed automaton with two clocks $x$ and $y$. Consider a state $(l, v)$ of  $T(A)$  with 
$(v(x), v(y))=(0.3,1.6)$. Consider an edge $e=(l, x < 1 \wedge y > 2, a, \{y\},l')$. 
Starting from $(l, (0.3,1.6))$, here is a sequence of transitions in $T(A)$ : 
$(l, (0.3,1.6)) \stackrel{0.5} {\longrightarrow}(l, (0.8, 2.1)) \stackrel{a\delete{e}}{\longrightarrow} (l', (0.8, 0))$.

For simplicity, we do not consider annotating 
locations with clock constraints (known as \emph{invariant conditions} \cite{HNSY1}). Our results extend in a
straightforward manner to timed automata with invariant conditions.
We now define various concepts that will be used in the paper.
\begin{definition} 
Let $A=(L, Act, l_0, E, C)$ be a timed automaton, and $T(A)$ be the TLTS corresponding to $A$. 
\begin{enumerate}
\item \textbf{Timed trace}: A sequence of delays and visible actions $d_1 a_1 d_2 a_2 \dots d_n a_n$
is called a timed trace iff there is a sequence of transitions 
$p_0 \to{d_1} p_1 \to{a_1} p_1'\to{d_1}p_2\to{a_2}p_2'\cdots \to{d_n}p_n\to{a_n}p'$ in $T(A)$, with 
$p_0\delete{=(l_0, v_0)}$ being a state of the timed automaton.
For a timed trace $tr = d_1 a_1 d_2 a_2 \dots d_n a_n$, $untime(tr) = a_1 a_2 \dots a_n$ represents the sequence of visible actions in $tr$.
\item \textbf{Zone}: A zone $z$ is a set of valuations $\{v \in \realpos^{|C|} \mid v \models \gamma\}$, where 
$\gamma$ is of the form $\gamma ::= \; x \smile c \:|\: x-y \smile c \:|\: g \wedge g$,
and $c \in \mathbb{Z}$, $x,y \in C$ and $\smile \: \in \: \{\le, <, =, >, \ge\}$.
$z \uparrow$ \delete{and $z \downarrow$ respectively}denotes the future \delete{and the past}of the zone $z$.
$z\uparrow=\{v+d \mid v \in z, d \geq 0\}$ is the set of all valuations 
reachable from $z$ by time elapse. 
\item \textbf{Pre-stability}: A zone $z_1$ is pre-stable with respect to another zone 
$z_2$ if $z_1 \subseteq preds(z_2)$ or $z_1 \cap preds(z_2) = \emptyset$ where
$preds(z)  \stackrel{def}{=} \{v \in \realpos^{|C|} \:|\: \exists v' \in z$ such that $v \to{\alpha} v'$,
$\alpha \in Act \cup \realpos$\}\delete{, $V$ being the possible set of clock valuations}.
\item \textbf{Canonical decomposition}: Let $z$ be a zone, and let $g =\bigwedge_{i=1}^n g_i \in \mathcal{B}(C)$, where 
each $g_i$ is of the form $x_i \smile c_i$. 
A canonical decomposition of $z$ with respect to  $g$ 
is obtained by splitting $z$ into a set of zones $z_1, \dots, z_m$ such that 
for each $1 \leq i \leq m$, and $1 \leq j \leq n$, \delete{(i)}for every valuation $v \in z_i$,\delete{is such that} either (i) $v \models g_j$, or (ii)
\delete{every valuation $v \in z_i$ is such that}$v \nvDash g_j$.

For example, consider the zone $z=x \geq 0 \wedge y \geq 0$ and the guard  $x \le 2 \wedge y > 1$. 
$z$ is split with respect to $ x \le 2$, and then with respect to $y > 1$, hence 
into four zones : 
$x \le 2 \wedge y \le 1$, $x > 2 \wedge y \le 1$, $x \le 2 \wedge y > 1$ and $x > 2 \wedge y > 1$.
\end{enumerate}
\end{definition}
Given a timed automaton $A$,\delete{the} a \emph{zone graph} of $A$ 
is used to check reachability in $A$. 
A \emph{node} in the zone graph is a pair consisting of a location and a zone. The edges between nodes  
are defined as follows.
$(l, z) \stackrel{a}{\rightarrow} (l', z')$, where $a \in Act$, if for every $v$  in
$z$, $\exists v'$ in $z'$ such that $(l, v) \stackrel{a}{\rightarrow}(l', v')$. If the 
zones corresponding to $(l, v)$ and $(l, v')$ are $z$ and $z'$ respectively and there is a transition 
in $T(A)$ such that $(l, v) \to{d} (l, v')$, then we have an edge $(l, z) \to{\varepsilon} (l, z')$ in the zone graph.
Every node has an $\varepsilon$ transition to itself and the $\varepsilon$ transitions are also transitive.
The zone $z'$ is called a \emph{delay successor zone} of zone $z$. Since $\varepsilon$ is reflexive, delay successor is also a reflexive relation.
For both $a$ and $\varepsilon$ transitions, if $z$ is a zone then $z'$ is also a zone, i.e. $z'$ is a convex set. A zone graph may be formally defined as a quadruple $(S, s_0, Lep, \rightarrow)$, where $S$ is the set 
of nodes of the zone graph, $s_0$ is the initial node, $Lep = Act \cup \{\varepsilon\}$ and $\rightarrow$ denotes the
set of transitions. 
$Z_{(A,p)}$  denotes a zone graph corresponding to the state $p$, i.e. the initial state of $Z_{(A,p)}$ is $p$.
\delete{A zone valuation graph $Z_{(A,p)}$ corresponds to a particular state $p$ of 
 the timed automaton $A$. The clock valuation of $p$ is same as the initial clock valuation corresponding 
 to which the zone valuation graph is created.}For a state $q \in T(A)$, $\mathcal{N}(q)$ represents the node
 of the zone \delete{valuation}graph with the same location as that of $q$ such that the zone corresponding to $\mathcal{N}(q)$ includes 
 the valuation of $q$. We often say that a state $q$ is in node $s$ to indicate that $q$ is in the zone associated with node $s$. For two zone \delete{valuation}graphs, 
 $Z_{(A_1,p)} = (S_1,s_p,Lep,\rightarrow_1)$, $Z_{(A_2,q)} = (S_2,s_q,Lep,\rightarrow_2)$ and a relation
 $\mathcal{R}\subseteq S_1\times S_2$, $Z_{(A_1,p)}\:\mathcal{R}\:Z_{(A_2,q)}$ iff $(s_p,s_q)\in \mathcal{R}$. 
 While checking $\mathcal{R}$, $\varepsilon$ is considered visible similar to an action in $Act$.
 An $\varepsilon$ action represents a delay $d \in \mathbb{R}_{\ge 0}$.
\begin{algorithm}
\caption{Construction of Zone Graph\newline 
\emph{Input: Timed automaton $A$ \newline
Output: Zone  graph corresponding to $A$}
}\label{algo-zonegraph}
\begin{algorithmic}[1]
{\small
\State Calculate $max_x^l$ for each location $l \in L$ and each clock $x \in C$. This is required 
for abstraction to ensure finite number of zones in the zone graph.
\State Initialize $Q$ to an empty queue.
\State $Enqueue(Q, <l_0, \emptyset>)$. \Comment{\textsf{\tiny Every element is a pair consisting of a location and its parent}}
\State $successors\_added = false$. \Comment{\textsf{\tiny flag set to true whenever  successors of a location are added to $Q$}}
\While {Q not empty}
\State $<l, l_p> = dequeue(Q)$
\If {$l_p\neq \emptyset$}, 
\State For the edge $l_p \to{g, a, X'} l$ in $A$, for each existing zone $z_{l_p}$ of $l_p$, create
the zone $z = (z_{l_p} \uparrow \cap \:g_{[X' \leftarrow \overline{0}]})$ of $l$, when $z \neq \emptyset$.
\State Abstract each of the newly created zones if necessary and for any newly created zone $z$, for location $l$,
if $\exists z_1$ of same location such that $z \cap z_1 \neq \emptyset$, then merge $z$ and $z_1$.
\State Update edges from zones of $l_p$ to zones of $l$ appropriately.
\State If a new zone of $l$ is added or an existing zone of $l$ is modified, then for all successors $l_j$ of $l$,
enqueue $<l_j, l>$ to Q.
\State $successors\_added := true$.
\EndIf
\State $new\_zone\_l := true$. \Comment{\textsf{\tiny flag set to false when the canonical decomposition does not produce further zones}}
\While {$new\_zone\_l$}
\State Split the existing zones $z$ of $l$ based on the canonical decomposition of the guards on the
outgoing edges of $l$ \Comment{\textsf{\tiny It is not always necessary for a split to happen.}}
\State For every zone $z$ of $l$, consider $z \uparrow$ and split it further based on the canonical
decomposition of the guards on the outgoing edges of $l$
\Comment{\textsf{\tiny Note that the zones created from this split are convex.}}
\State Abstract each of the newly created zones if necessary and update edges appropriately.
\State If new zones are not created then set $new\_zone\_l$ to $false$.
\EndWhile
\If {any new zones of $l$ are created or any existing zones of $l$ are modified due to the canonical
decomposition of the outgoing edges of $l$ and $successors\_added = false$} 
\State for all the successor locations $l_j$ of $l$ to Q, enqueue $<l_j, l>$ to Q.
\EndIf
\EndWhile
\State  /* \textsf{Phase 2 :  In this phase, pre-stability is enforced} */
\State $new\_zone = true$
\While {$new\_zone$}
\State $new\_zone  = false$
\ForAll {edges $l_i \to{g, a, X'} l_j$}
\ForAll {pairs of zones $z_{lik}$, $z_{ljm}$ such that $z_{lik} \to{\alpha}z_{ljm}$ is an edge in the 
zone graph where $\alpha \in Act \cup \realpos$}
\If {$z_{lik}$ is not pre-stable with respect to $z_{ljm}$,}
Split $z_{lik}$ to make it pre-stable with respect to $z_{ljm}$.   \Comment{\textsf{\tiny Note that this split still maintains 
convexity of $z_{lik}$ since the zone is split entirely along an axis that is parallel to the diagonal in 
the $|C|$-dimensional space.}}
\State $new\_zone := true$
\State Update the edges
\EndIf
\EndFor
\EndFor
\EndWhile
}
\end{algorithmic}
\end{algorithm}
The detailed algorithm for creating the zone graph has been described in algorithm \ref{algo-zonegraph}
and consists of two phases, the first one being a forward analysis of the timed automaton while the second phase
ensures pre-stability in the zone graph. The set of valuations for every 
location is initially split into zones based on the canonical decomposition of its outgoing transition\delete{as discussed in
\cite{TY1}}. The forward analysis may cause a zone graph to become infinite \cite{BBFL1}. Several kinds of abstractions
have been proposed in the literature \cite{DT1}\cite{BBFL1}\cite{BBLP1}. We use \emph{location dependent maximal constants}
abstraction \cite{BBFL1} \delete{to be used for creating}to ensure finiteness of the \delete{finite}zone \delete{valuation}graph. In algorithm \ref{algo-zonegraph}, $max_x^l$ denotes the maximum constant in location $l$ beyond which the value of clock $x$ is irrelevant.
After phase 2, pre-stability ensures the following: For a node $(l, z)$ in the zone graph, with $v \in z$,
\delete{such that for a $v \models z$,}for a timed trace $tr$,
if $(l, v) \to{tr} (l'', v'')$, with $v'' \in z''$, then 
$\forall v' \; \delete{\models}\mbox{in}\; z$, $\exists tr'.(l, v') \to{tr'} (l'', \tilde{v})$, with
$untime(tr') = untime(tr)$ and $\tilde{v} \in z''$. 
\delete{Here $untime(tr)$ represents the sequence of visible actions in $tr$.}According to the construction given
in algorithm \ref{algo-zonegraph}, for a particular location of the timed automaton, the zones corresponding 
to any two nodes are disjoint. Convexity of the zones and pre-stability property \delete{of the zones}together ensure
that a zone with elapse of time is intercepted by a single hyperplane of the form $x = h$ as in the case of regions,
where $x \in C$ and $h \in \mathbb{N}$\delete{in the future}. Some approaches for preserving convexity and implementing pre-stability have been discussed \delete{in detail}in \cite{TY1}. 
\begin{figure}[h]
\centering
\vspace{-10pt}
\includegraphics[width=0.4\textwidth]{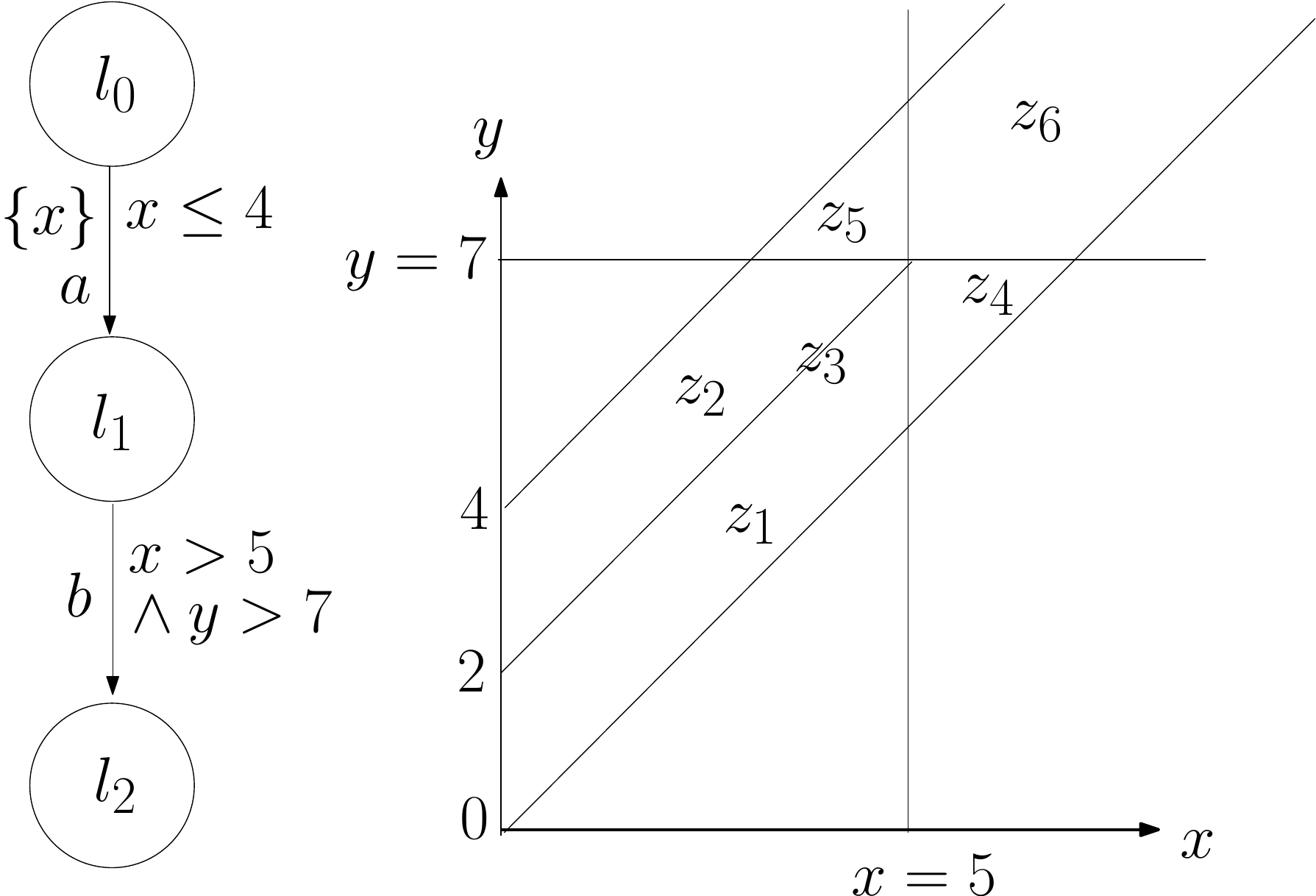}
\caption{\label{fig-delayZone} A timed automaton and the zones for location $l_1$}
\vspace{-10pt}
\end{figure}
As an example consider the timed automaton in Figure \ref{fig-delayZone}. The zones corresponding to location $l_1$ as produced through algorithm \ref{algo-zonegraph} are shown in the right side of the figure.

A similar construction of zone graph has also been used in \cite{GNA1}. In the construction used in \cite{GNA1}, in the final phase, the nodes corresponding to a particular location with zones that are time abstracted bisimilar to each other are merged as long as the merged zone is convex. Though this may reduce the number of zones in the final zone graph, the operation itself is exponential in the number of clocks of the timed automaton. Due to the absence of this merging phase in the algorithm described in this paper, while checking the existence of the relations following the method described here, one may need to consider more pairs of \delete{corner points}states, but we expect this overhead to be less compared to the expensive operation of merging the nodes with time abstracted bisimilar zones.

\section{Equivalences for Timed Systems}\label{sec-timedrels}
In this section, we define the timed and the time abstracted
relations considered in this work. We only consider the strong form of these relations here. We enumerate a few clauses first using which we define $p_1 \:\mathcal{R}\: p_2$  where $p_1$ and $p_2$ are two timed automata states and $\mathcal{R}$ is a timed or a time abstracted relation.
\begin{enumerate}
\item \label{rel-Act}$\forall a \in Act \wedge \forall p_1'$, $p_1 \stackrel{a}{\rightarrow} p_1' \Rightarrow [\:\exists p_2' : p_2 \stackrel{a}{\rightarrow} p_2' \wedge p_1' \mathcal{R} p_2'\:]$
\vspace{-5pt}
\item \label{rel-Act2}$\forall a \in Act \wedge \forall p_2'$, $p_2 \stackrel{a}{\rightarrow} p_2' \Rightarrow [\:\exists p_1' : p_1 \stackrel{a}{\rightarrow} p_1' \wedge p_1' \mathcal{R} p_2'\:]$
\vspace{-5pt}
\item \label{rel-delayAct}$\forall a \in Act \wedge \forall p_1'$, $p_1 \stackrel{a}{\rightarrow} p_1' \Rightarrow [\:\exists p_2'\: \exists d \in \mathbb{R}_{\ge 0} : p_2 \stackrel{d}{\rightarrow} \stackrel{a}{\rightarrow} p_2' \wedge p_1' \mathcal{R} p_2'\:]$
\vspace{-5pt}
\item \label{rel-delayActDelay}$\forall a \in Act \wedge \forall p_1'$, $p_1 \stackrel{a}{\rightarrow} p_1' \Rightarrow [\:\exists p_2'\: \exists d_1, d_2 \in \mathbb{R}_{\ge 0} : p_2 \stackrel{d_1}{\rightarrow} \stackrel{a}{\rightarrow} \stackrel{d_2}{\rightarrow} p_2' \wedge p_1' \mathcal{R} p_2'\:]$
\vspace{-5pt}
\item \label{rel-matchDelay}$\forall d \in \mathbb{R}_{\ge 0} \wedge \forall p_1'$, $p_1 \stackrel{d}{\rightarrow} p_1' \Rightarrow [\:\exists p_2' : p_2 \stackrel{d}{\rightarrow} p_2' \wedge p_1' \mathcal{R} p_2'\:]$
\vspace{-5pt}
\item \label{rel-delay1Delay2}$\forall d \in \mathbb{R}_{\ge 0} \wedge \forall p_1'$, $p_1 \stackrel{d}{\rightarrow} p_1' \Rightarrow [\:\exists p_2' \: \exists d' \in \mathbb{R}_{\ge 0} : p_2 \stackrel{d'}{\rightarrow} p_2' \wedge p_1' \mathcal{R} p_2'\:]$
\vspace{-5pt}
\item \label{rel-delayLessDelay}$\forall d \in \mathbb{R}_{\ge 0} \wedge \forall p_1'$, $p_1 \stackrel{d}{\rightarrow} p_1' \Rightarrow [\:\exists p_2' \: \exists d' \in \mathbb{R}_{\ge 0}\: \wedge \: d \le d' \: : p_2 \stackrel{d'}{\rightarrow} p_2' \wedge p_1' \mathcal{R} p_2'\:]$
\vspace{-5pt}
\item \label{rel-delayMoreDelay}$\forall d \in \mathbb{R}_{\ge 0}\wedge \forall p_2'$, $p_2 \stackrel{d}{\rightarrow} p_2' \Rightarrow [\:\exists p_1' \: \exists d' \in \mathbb{R}_{\ge 0}\: \wedge \: d \ge d' \: : p_1 \stackrel{d'}{\rightarrow} p_1' \wedge p_1' \mathcal{R} p_2'\:]$
\end{enumerate}
$\mathcal{R}$ is a timed simulation if the clauses \ref{rel-Act} and \ref{rel-matchDelay} hold. For each $(p_1, p_2) \in \mathcal{R}$, $p_2$ time simulates $p_1$. $\mathcal{R}$ is a \emph{timed simulation equivalence} if $p_1$ time simulates $p_2$ and $p_2$ time simulates $p_1$. A symmetric timed simulation is a \emph{timed bisimulation} relation. A symmetric relation that satisfies clauses \ref{rel-Act} and \ref{rel-delay1Delay2} is a \emph{time abstracted bisimulation}. A relation that is symmetric and satisfies clauses \ref{rel-delayAct} and \ref{rel-delay1Delay2} is a \emph{time abstracted delay bisimulation relation}. A symmetric relation satisfying clauses \ref{rel-delayActDelay} and \ref{rel-delay1Delay2} is a \emph{time abstracted observational bisimulation}. A \emph{timed performance prebisimulation relation} \cite{GNA1} satisfies the clauses \ref{rel-Act}, \ref{rel-Act2}, \ref{rel-delayLessDelay} and \ref{rel-delayMoreDelay}.

The corresponding largest bisimulation relations are called \emph{bisimilarity} relations and they are \emph{timed bisimilarity} ($\sim_t$), \emph{time abstracted bisimilarity} ($\sim_u$), \emph{time abstracted delay bisimilarity} ($\sim_y$), \emph{time abstracted observational bisimilarity} ($\sim_o$) whereas the largest prebisimulation relation is called \emph{timed performance prebisimilarity} ($\precsim$). $p \precsim q$ denotes that $p$ is \emph{at least as fast as} $q$. It is easy to see from the definitions that timed bisimilarity implies time-abstracted bisimulation whereas the converse is not true. Besides, the definitions imply $\sim_u \:\subseteq\: \sim_y \:\subseteq \:\sim_o$. Also the existence of a bisimulation relation between two states implies the existence of the corresponding simulation equivalence and timed performance prebisimilarity lies in between timed bisimulation and time abstracted bisimulation. Hence we have $\sim_t \:\subseteq \:\precsim \:\subseteq \:\sim_u \:\subseteq \:\sim_y \:\subseteq \:\sim_o$ and similar containment 
relations also exist among the corresponding simulation equivalences.
\section{Deciding Relations for Timed Automata}\label{sec-dec}
In this section, we present a unifying approach to decide several relations for timed automata using the zone graph constructed in algorithm \ref{algo-zonegraph}.
\subsection{Deciding Timed Bisimulation}
Timed bisimulation has been proven to be decidable for timed automata \cite{KC1}. A product construction technique on the region graphs has been used in \cite{KC1} whereas  in \cite{WL1}, a product construction is applied on zone graphs instead for deciding timed bisimulation. Though decidable, 
timed bisimulation may have uncountably many equivalence classes\cite{ACH1}.
We define \emph{corner point bisimulation} relation and show that 
corner point bisimulation coincides with timed bisimulation. With corner point bisimulation, only a finite
number of pairs of corner points are needed for bisimilarity checking. Further,
our method eliminates the product construction on zone graphs.  

Let $A$, $B$ be timed automata having $C_A, C_B$ as the respective maximum constants used in the constraints appearing in the two automata. Let 
 $p$ and $q$ be two states in $T(A)$ and $T(B)$ respectively. We show that 
\begin{enumerate}[(i)]
\item if $p$ and $q$ are initial states, or states where \emph{all} clock valuations are integers, then timed bisimulation for $p, q$ can be decided by checking delays of the form $n$, $n + \
\delta$ or $n - \delta$, where $n \in \{0, 1,  \dots, C\}$, $C = max(C_A, C_B)$, and $\delta$ is a symbolic value for an infinitesimal positive quantity.
\vspace{-5pt}
\item If there is some clock $y$ having a non-zero rational fractional part, then along with the delays of the form mentioned above, we check delays of the form $f$, $f + \delta$ or $f - \delta$, with $f  = 1 - frac(v(y))$, $frac(v(y))$ is the fractional part of the value of clock $y$.
\end{enumerate}
Delays of the form mentioned above are called \emph{corner point delays} or \emph{cp-delays}.
We define corner point bisimulation formally below.
\begin{definition} \label{def-cpsim}\hfill
\vspace{-5pt}
\begin{enumerate}
 \item Corner point simulation (cp-simulation): A relation $\mathcal{R}$ is a corner point simulation relation, if for every pair of timed automata states $(p, q) \in \mathcal{R}$, the following conditions hold. \begin{enumerate}[(i)]
\vspace{-5pt}
\item For every visible action $a \in Act$, if $p \to{a} p'$, then $\exists q'$ such that $q \to{a}q'$ and $p' \mathcal{R}q'$
\vspace{-5pt}
\item Considering the maximum possible delay $d$ from $p$, if $p \to{d}p'$ and $p'$ is in node $\mathcal{N}(p)$, then $\exists q'$ such that $q \to{d}q'$ and $p' \mathcal{R} q'$
\vspace{-5pt}
\item For every node $\mathcal{N}(p') \neq \mathcal{N}(p)$ such that $\mathcal{N}(p) \to{\varepsilon} \mathcal{N}(p')$, considering the minimum delay $d$ from $p$, if $p \to{d}p'$, then $\exists q'$ such that $q \to{d}q'$ and $p' \mathcal{R} q'$
\end{enumerate}
Here $q$ cp-simulates $p$. A symmetric corner point simulation relation is a corner point bisimulation (cp-bisimulation).
\item Corner point trace:  A timed trace from a state $p$ to $p'$, where all the delays are cp-delays 
is called a corner point trace.
\end{enumerate}
\end{definition}
\begin{lemma} \label{lem-cpfinite}
For checking whether the timed automata states $p$ and $q$ are related through corner point simulation or corner point bisimulation relation, there are only finitely many pairs of states that need to be considered.
\end{lemma}
This is due to the fact that for any $(p, q) \in \mathcal{R}$, $\mathcal{R}$ being a cp-bisimulation relation, the valuations of all the clocks appearing in both $p$ and $q$ are of the form $n$, $n+ \delta$ or $n-\delta$, where $n \in \{0, 1, \dots, C_A\}$ or $n \in \{0, 1, \dots, C_B\}$. If $p \in T(A)$ and $q \in T(B)$, then $C_A$ and $C_B$ are the maximum constants appearing in $A$ and $B$ respectively.

\begin{theorem} \label{thm-cp}
Corner point simulation and corner point bisimulation relations are decidable.
\end{theorem}

\delete{
\begin{lemma} \label{lem-tbimpcpbisim}
For two timed automata states $p$ and $q$, $p \sim_t q \Rightarrow p \mathcal{R} q$, where $\mathcal{R}$ is a corner point bisimulation relation.
\end{lemma}

We will now prove that corner point bisimulation between two timed automata states implies that they are timed bisimilar.
\begin{fact}
If $p$ and $q$ are not timed bisimilar, then one of the following conditions hold true.
\begin{itemize}
\item There exists a timed trace $tr$ such that $p \to{tr} p'$ and $\forall q'$ such that $q \to{tr} q'$ and $sort(p') \neq sort(q')$.
\item There exists a timed trace $tr$ such that $q \to{tr} q'$ and $\forall p'$ such that $p \to{tr} p'$ and $sort(q') \neq sort(p')$.
\end{itemize}
\end{fact}
Taking contrapositive of the above fact, if for every timed trace $tr$ from $p$, such that $p \to{tr} p'$, $\exists q'$ such that $q \to{tr} q'$ and $sort(p') = sort(q')$ and similarly for every timed trace $tr$ from $q$, such that $q \to{tr} q'$, there exists a $p'$ such that $p \to{tr} p'$ and $sort(q') = sort(p')$, then $p$ and $q$ are timed bisimilar.

We prove that if $p$ and $q$ are cp-bisimilar, then the above condition holds. We state this formally below. In the statement of the lemma, we abuse the notation $p_1 + d_1$ to denote a state where all clock valuations of $p_1$ have been increased by $d_1$.

\begin{lemma} \label{lem-cpbisimImp}
If $p_1$ and $q_1$ are cp-bisimilar, then the following conditions hold true for all $n \in \mathbb{N}$.
\begin{itemize}
\item For all delays $d_1, d_2, \dots, d_n \in \mathbb{R}_{\ge 0}$ and $\forall a_1 \in sort(p_1 + d_1)$, $\forall a_2 \in sort(p_2 + d_2), \dots$, $\forall a_n \in sort(p_n + d_n)$ such that
	$p_1 \to{d_1} \to{a_1} p_2 \to{d_2}\to{a_2} \cdots p_n \to{d_n}\to{a_n} p_{n+1}$, 
	$\exists a_1 \in sort(q_1 + d_1)$, $\exists a_2 \in sort(q_2 + d_2)$, \dots, $\exists a_n \in sort(q_n + d_n)$, such that
	$q_1 \to{d_1}\to{a_1}q_2 \to{d_2}\to{a_2} \cdots q_n \to{d_n}\to{a_n}q_{n+1}$ and $p_{n+1}$ and $q_{n+1}$ can perform the same set of actions after same delay $d_{n+1} \in \realpos$, i.e. $sort(p_{n+1} + d_{n+1}) = sort(q_{n+1} + d_{n+1})$.
\item For all delays $d_1, d_2, \dots, d_n \in \mathbb{R}_{\ge 0}$ and $\forall a_1 \in sort(q_1 + d_1)$, $\forall a_2 \in sort(q_2 + d_2), \dots$, $\forall a_n \in sort(q_n + d_n)$ such that 
	$q_1 \to{d_1} \to{a_1} q_2 \to{d_2}\to{a_2} \cdots q_n \to{d_n}\to{a_n} q_{n+1}$, 
	$\exists a_1 \in sort(p_1 + d_1)$, $\exists a_2 \in sort(p_2 + d_2)$, \dots, $\exists a_n \in sort(p_n + d_n)$, such that
	$p_1 \to{d_1}\to{a_1}p_2 \to{d_2}\to{a_2} \cdots p_n \to{d_n}\to{a_n}p_{n+1}$ and $p_{n+1}$ and $q_{n+1}$ can perform the same set of actions after same delay $d_{n+1} \in \realpos$, i.e. $sort(p_{n+1} + d_{n+1}) = sort(q_{n+1} + d_{n+1})$.
\item If $p \to{tr} p'$ and $q \to{tr}q'$ where $\mathcal{N}_{p'}$ and $\mathcal{N}_{q'}$ are the nodes containing timed states $p'$ and $q'$ respectively, then there exists at least one corner point trace $tr_i$ such that $untime(tr) = untime(tr_i)$ and $p \to{tr_i} \tilde{p}$ and $q \to{tr_i} \tilde{q}$ such that $\tilde{p}$ is a state in the node $\mathcal{N}_{p'}$ and $\tilde{q}$ is a state in the node $\mathcal{N}_{q'}$. $untime(tr)$ is obtained by removing 
all the delays from the timed trace $tr$.
\end{itemize}
\end{lemma}

\begin{lemma} \label{lem-cpImpTB}
For two timed automata states $p$ and $q$, $p \mathcal{R} q \implies p \sim_t q$, where $\mathcal{R}$ is a corner point bisimulation relation.
\end{lemma}

We state here the main theorem of the paper for deciding timed bisimulation by deciding cp-bisimulation by combining Lemma \ref{lem-tbimpcpbisim} and \ref{lem-cpImpTB}.
} 

\begin{theorem} \label{thm-timedbisim}
For two timed automata states $p$ and $q$, 
\begin{enumerate}
\item \label{lem-tbimpcpbisim} $p \sim_t q \Rightarrow p \mathcal{R} q$, where $\mathcal{R}$ is a corner point bisimulation relation.
\item \label{lem-cpImpTB} $p \mathcal{R} q \Rightarrow p \sim_t q$, where $\mathcal{R}$ is a corner point bisimulation relation.
\item $p$ and $q$ are timed bisimilar if and only if $p$ and $q$ are cp-bisimilar.
\end{enumerate}
\end{theorem}

Theorem \ref{thm-timedbisim} shows that the decidability of cp-bisimulation is sufficient for timed bisimulation.

\\[.2cm]{\bf Synthesis of Distinguishing Formulae.} 
Given two timed automata $A$ and $B$ which are not timed bisimilar, 
we propose a technique that synthesizes a formula that captures 
the differences between $A$ and $B$. 

In \cite{LLW1}, a \emph{characteristic formula} for timed automata has been defined using a certain fragment of the $\mu$-calculus presented in \cite{HNSY1}. Timed bisimilarity between two timed automata is decided by comparing one timed automaton with the characteristic formula of the other. A characteristic formula is a significantly complex formula describing the entire behaviour of the timed automaton. Here we describe how we can in general generate a simpler formula using a fragment of the logic described in \cite{LLW1}. The logic we use for generating the distinguishing formula has been described in \cite{LAKJ1} which is a timed extension of Hennessy-Milner logic and does not contain any recursion as opposed to the logic used in \cite{LLW1}. The set $\mathcal{M}_t$ of Hennessy-Milner logic formulae with time over a set of actions $Act$, set $D$ of formula clocks (distinct from the clocks of any timed automaton) is generated by the abstract syntax

\begin{center}
$\phi ::= \true \;|\; \false \;|\; \phi \wedge \psi \;|\; \phi \vee \psi \;|\; \langle a \rangle \phi \;|\; [a] \phi \;|\; \existsdelay \phi \;|\; \foralldelay \phi \;|\; x \isin \phi \;|\; g$ \\
\end{center}
where $a \in Act$, $x \in D$ and $g \in \mathcal{B}(D)$.
The logic used in \cite{GL1} for constructing distinguishing formula uses an explicit negation rather than using the operators $[ \; ]$ and $\foralldelay$. Besides a distinguishing formula in \cite{GL1} uses real delays whereas in our case, the formula clock values are compared with integers. Also the distinguishing formula synthesized in \cite{GL1} considers the entire branching structure of the given automata whereas in our case, the formula is synthesized from the moves in a game and is thus more succinct.

Given a timed automaton $A$, $\mathcal{M}_t$ is interpreted over an extended state $\langle (l, v) u\rangle$, where $(l, v)$ is a state of $A$ and $u$ is a time assignment of $D$. Transitions between the extended states are defined by: $\langle (l, v) u\rangle \to{d}\langle (l, v+d) u+d\rangle$ and $\langle (l, v) u\rangle \to{a} \langle (l', v') u'\rangle$ iff $\langle (l, v) \rangle \to{a} \langle (l', v')\rangle$ and $u = u'$.

$\existsdelay \phi$ holds in an extended state if there \emph{exists} a delay transition leading to an extended state satisfying $\phi$. Similarly $\foralldelay$ denotes universal quantification over delay transitions, and $\langle a \rangle$ and $[a]$ respectively denote existential and universal quantification over $a$-transitions. The formula $x \isin \phi$ introduces a formula clock $x$ and initializes it to 0, i.e. $\langle (l, v) u\rangle \models x \isin \phi \implies \langle (l, v) u_{[x \leftarrow \bar{0}]}\rangle \models \phi$. The formula clocks are used in formulas of the \delete{type}form $g$ which is satisfied by an extended state if the values of the formula clocks used in $g$ satisfy the specified relationship. A formula is said to be \emph{closed} if each occurrence of a formula clock $x$ is within the scope of an $x \isin$ construct.

While checking cp-bisimulation, we describe below a method to generate a closed formula in $\mathcal{M}_t$ that distinguishes two timed automata that are not timed bisimilar. For constructing the formula, we consider a cp-bisimulation game between the \emph{initial states} of the timed automata
 which can be thought of as a bisimulation game (see \cite{CS2}) for deciding timed bisimulation between \delete{the initial states of}two timed automata. 
The game is played between two players, the challenger and the defender. Each round of the game consists of
the challenger choosing one of the zone graphs and making a move as defined in the definition of the cp-bisimulation relation. The defender tries to replicate the move in the other zone graph. The defender loses the game if after a finite sequence of rounds, the challenger makes a move on one zone graph which the defender cannot replicate on the other. If the defender loses the game, we look at the sequence of moves chosen by the challenger to construct the distinguishing formula as follows:

Given two timed automata $A$ and $B$ (with $C_1 \cap C_2 = \emptyset$ where $C_1$ and $C_2$ are clocks of $A$ and $B$ respectively) and their zone graphs being $Z_{A}$ and $Z_{B}$, let us suppose without loss of generality that the challenger makes a move on $Z_{A}$ in the first round. We derive a formula from the moves of the game which is satisfied by automaton $A$ and not by automaton $B$. The set of formula clocks are disjoint from the clocks in $A$ and $B$. The distinguishing formula $\zeta$ is initialized to $x_1 \:\isin\:()$ with the introduction of a formula clock $x_1$. Corresponding to every clock $y$ in $C_1 \cup C_2$, there exists a clock $x \in D$ such that their valuations are the same, i.e. $v(y) = v(x)$. We can define a mapping $\eta: C_1 \cup C_2 \rightarrow D$. With the introduction of the formula clock $x_1$ mentioned above, we have $\forall y \in C_1 \cup C_2$, $\eta(y) = x_1$. Whenever one or more clocks are reset either in $A$ or $B$ corresponding to the visible actions chosen by the challenger and the 
defender, a new formula clock is introduced. If $U \subseteq C_1$ and $V \subseteq C_2$ be the subset of clocks reset for actions chosen in a particular round, then a new formula clock $x_i$ is introduced such that $\forall y \in U \cup V, \eta(y) = x_i$. Subformulas of $\zeta$ are always added to the scope of the innermost or the last added clock in $\zeta$. The subformulas of $\zeta$ are added based on the move of the challenger as described below.
\begin{itemize}
\item[-] The challenger performs action $a \in Act$ in $Z_{A} (Z_{B})$. If the defender can reply with $a$ in $Z_{B} (Z_{A})$ and in either of the moves at least one clock is reset in the corresponding timed automata, then add $\langle a \rangle \; x \; \isin \; (\;) \;(\:[a] \; x \; \isin \; (\;)\:)$ to the innermost scope of $\zeta$, where $x$ is a new formula clock. If no clock is being reset, then simply add $\langle a \rangle ([a])$ to the innermost scope of $\zeta$. If the defender cannot reply to the move of the challenger, then append $\langle a \rangle \true \;(\:[a]\false\:)$ in the innermost scope and declare $\zeta$ to be the distinguishing formula.
\item[-] The challenger performs a delay $d$ in $Z_{A} (Z_{B})$, where the delays are as defined in the cp-bisimulation relation, i.e. cp-delays and it reaches a state with clock valuation $v$. For every clock $y$ in $Z_{A} (Z_{B})$, we construct a subformula as follows: if $v(y) = n$ is an integer, then construct the subformula $x = n$, where $x$ is $\eta(y)$. If $v(y)$ is of the form $n + \delta$, where $n \in \mathbb{N}$, then construct the subformula $n < x < n + 1$ and if $v(y)$ is of the form $n - \delta$, then construct the subformula $n-1 < x < n$. Conjunct all these subformulas to obtain $\psi$ and append $\existsdelay (\psi) \; (\:\foralldelay (\psi)\:)$ to the innermost scope of $\zeta$.
\end{itemize}
Note that in $\zeta$, the subformulas of the form $g$ define the smallest set of extended states reachable along the trace followed in $Z_{A}$ that can be specified by subformulas of the form $g$. 
\begin{figure}[t]
\centering
\includegraphics[width=.6\textwidth]{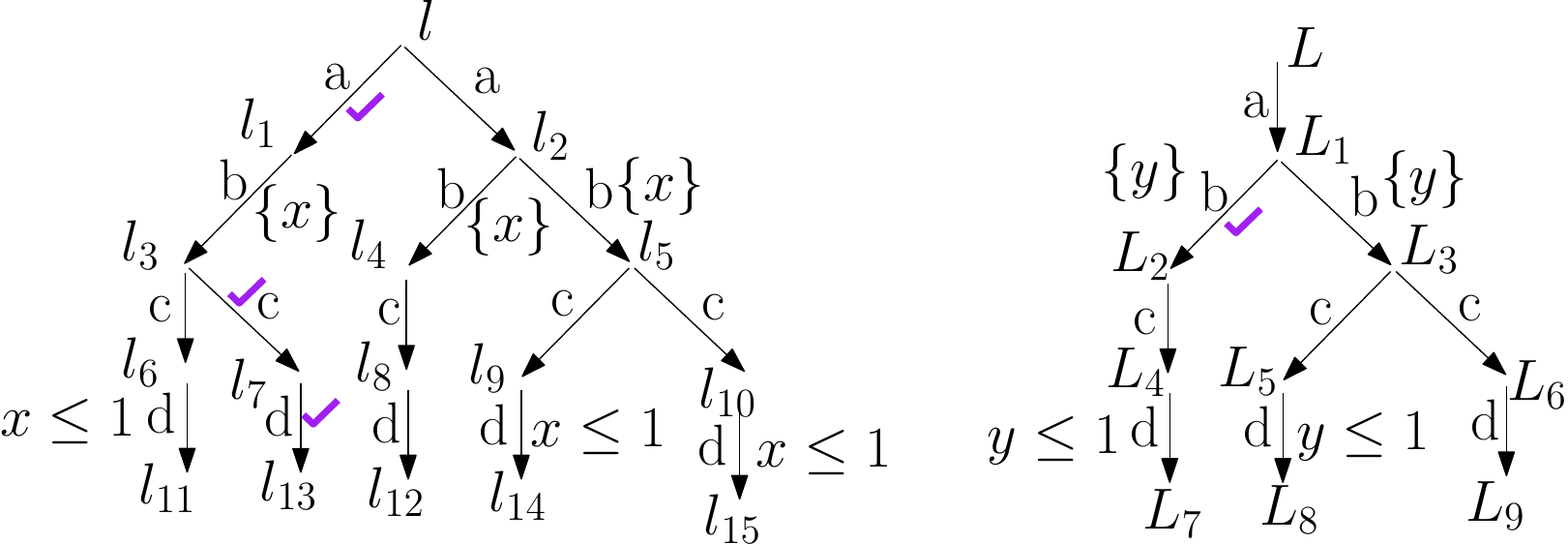}
\caption{\label{fig-formula} Example of distinguishing formula in $\mathcal{M}_t$}
\end{figure}
In the two timed automata shown in Figure \ref{fig-formula}, the moves of the challenger are marked by a $\checkmark$ mark. The challenger starts making a move from the automaton on the left and hence we construct a formula in $\mathcal{M}_t$ that is satisfied by the automaton on the left but not by the automaton on the right. Following the steps mentioned above, we obtain the formula $x_1 \;\isin\;(\langle a \rangle [b] \; x_2 \; \isin \;(\langle c \rangle \; \existsdelay (1 < x_2 < 2 \wedge \langle d \rangle \true)))$.

We note that though the cp-bisimulation relation can be decided between any two timed automata states with arbitrary clock valuation, the distinguishing formula is constructed only while checking the relation between the initial states of the two timed automata.

The technique described here to check for lack of timed bisimilarity can be 
adapted \delete{for}to many of the other relations studied in this paper. 
\delete{The full version of the paper, 
will contain the details; o}Our prototype tool implementation which is currently underway will 
also incorporate them.

\subsection{Deciding Timed Performance Prebisimulaton} \label{subsec-decPrebisim}
In this section, we  define \emph{corner point prebisimulation} relation and show that this relation coincides with the timed performance prebisimulation relation. We use the zone graph constructed according to algorithm 1  for checking 
corner point prebisimulation. Unlike the case of timed bisimulation, a product construction on zone graphs
is not useful for deciding timed performance prebisimulation relation : for example,  consider two simple timed automata with clocks $x$ and $y$ respectively, each with two locations and one edge between them such that the edge in one automaton is labelled with $\langle x = 2, a, \emptyset \rangle$ while the other is labelled with $\langle y = 5, a, \emptyset \rangle$. These two timed automata are timed performance prebisimilar though a product on the region graphs of these two automata does not produce an action transition and thus does not have sufficient information to show that one of the automata can actually perform action $a$ following a lesser delay. 

\begin{figure}
\centering
\includegraphics[width=0.6\textwidth]{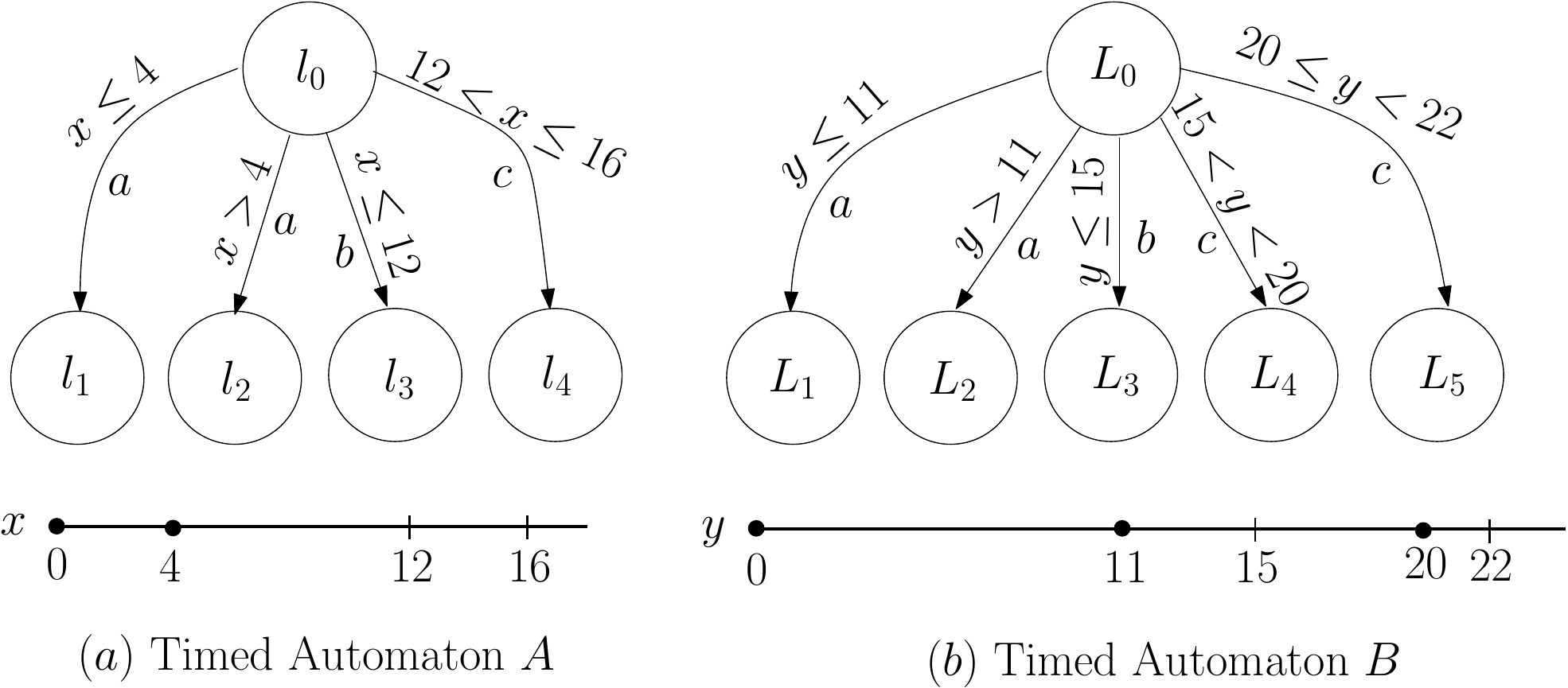}
\caption{\label{fig-cp_prebisimGame} Example of corner point prebisimulation game}
\end{figure}

We define the corner point prebisimulation relation in terms of a two player game that is similar to the bisimulation game\delete{since the operational definition is more tedious}.  The game is played between two players, challenger and defender
on the zone graphs (as constructed in algorithm 1) of two timed automata.  
In each round, the challenger chooses a side and the defender chooses the other side. After selecting a side, the challenger can either perform a visible action or a delay action. Note that in the corner point prebisimulation relation, 
the delays are cp-delays as given by Definition \ref{def-cpsim}. 
Two timed automata states $p$ and $q$ in $T(A)$ and $T(B)$ respectively\delete{$A, B$} are cp-prebisimilar, denoted \delete{$A \precsim_{cp} B$}$p \precsim_{cp} q$, if starting from $p$ and $q$, the defender wins and the cp-delay moves in $A$ are less than or equal to the corresponding cp-delay moves in $B$. \delete{For two timed automata, $A$ and $B$} We write $A \precsim_{cp} B$ if $p \precsim_{cp} q$, where $p$ and $q$ are respectively the initial states of $T(A)$ and $T(B)$.
We now explain the possible moves of the game on the respective zone graphs $Z_{(A, p)}$ and $Z_{(B, q)}$. Each move results in a new state in a possibly new zone from which the next move is made in the next round.
\begin{itemize}
\item[-] (Challenger chooses $T(A)$ (Move 1)): Performs a visible action $a \in Act$. \\
(Defender chooses $T(B)$): i) \delete{Defender replies with performing}Performs the same action $a$.

\item[-] (Challenger chooses $T(A)$ (Move 2)): Performs maximum delay $d$ and stays inside the same zone. \\
(Defender chooses $T(B)$): i) Performs delay $d$.
\item[-] (Challenger chooses $T(A)$ (Move 3)): Performs the minimum delay $d$ and moves to the next zone.\\
(Defender chooses $T(B)$ and performs one of the following delays):  i) delay $d$ or ii) cp-delays $d' \ge d$ that take $q$ to the delay successor zones.
\item[-] (Challenger chooses $T(B)$ (Move 1)): Performs a visible action $a \in Act$. \\
(Defender chooses $T(A)$): i) \delete{Defender replies with performing}Performs the same action $a$.
\item[-] (Challenger chooses $T(B)$ (Move 2)): Performs maximum delay $d$ and stays inside the same zone.\\
(Defender chooses $T(A)$ and performs one of the following delays): i) delay $d$ itself or ii) Consider cp-delay $d' \le d$, such that $p$ on elapsing $d'$ reaches the end of the same zone or other delay successor zones.
\item[-] (Challenger chooses $T(B)$ (Move 3)): Performs the minimum delay $d$ and moves to the next zone.\\
(Defender chooses $T(A)$ and performs one of the following delays):  i) delay $d$ itself or ii) cp-delays $d' \le d$ such that $p$ on elapsing $d'$ reaches the beginning of the delay successor zones or iii) cp-delays $d' \le d$ such that it reaches the end of the same or other delay successor zones.
\end{itemize}
Figure \ref{fig-cp_prebisimGame} illustrates how a corner point prebisimulation game is played between the challenger and the defender and shows \delete{the various moves of}each of the moves described above.
Note that in automata $A$ and $B$, for locations $l_0$ and $L_0$ respectively,
\begin{itemize}
\item[-] action $a$ is enabled at all delays.
\vspace{-5pt}
\item[-] actions $a$ and $b$, both are enabled when $x \le 12$ and $y \le 15$ respectively and
\vspace{-5pt}
\item[-] action $c$ is enabled in the interval $12 < x \le 16$ \delete{for $l_0$}in $A$ and in the interval $15 < y < 22$ \delete{for $L_0$}in $B$.
\end{itemize}
\vspace{-5pt}
For these two automata, we can see that $(l_0, x = 0) \precsim (l_1, y = 0)$, i.e. the automaton $A$ is at least as fast as automaton $B$. In $A$, the zones created using the algorithm \ref{algo-zonegraph}
corresponding to $l_0$ are $x \le 4$, $4 < x \le 12$, $12 < x \le 16$ and $x > 16$, whereas in $B$, the zones created are $y \le 11$, $11 < y \le 15$, $15 < y < 20$, $20 \le y < 22$ and $y \ge 22$. In Figure \ref{fig-cp_prebisimGame}, we also show a representative diagram of these zones. The dots on the axis of the clock denote the boundary of a zone that does not signify any change in behaviour whereas the small vertical lines on the axis of the clock denote an actual change in behaviour. A formal definition of \emph{similarity in behaviour} is given in the appendix.
\begin{table}[t]\footnotesize
\begin{tabular}{|c|m{5cm}|m{8cm}|} \hline
\multicolumn{3}{|c|}{Challenger Selects $Z_A$}  \\ \hline
& Challenger moves & Defender moves (on $Z_B$) \\ \hline
Move 1 & $(l_0, x = 0) \to{a} (l_1, x = 0)$ & $(L_0, y = 0) \to{a} (L_1, y = 0)$ \\ \hline
Move 2 & $(l_0, x = 0) \to{4} (l_0, x = 4)$ & $(L_0, y = 0) \to{4} (L_0, y = 4)$ \\ \hline
\multirow{2}{*}{Move 3} & $i) (l_0, x = 0) \to{4+\delta} (l_0, x = 4+\delta)$ & $(L_0, y = 0) \to{4+\delta} (L_0, y = 4+\delta)$ (Note that $(L_0, y = 11 + \delta)$ is not timed performance prebisimilar to $(l_0, x = 4 + \delta)$.) \\ \cline{2-3}
& $ii) (l_0, x = 4+\delta) \to{8} (l_0, x = 12 +\delta)$ & $(L_0, y = 4 + \delta) \to{11} (L_0, y =  15 + \delta)$ \\ \hline
\end{tabular}
\caption{Moves of the cp-prebisimulation game corresponding to automata shown in Figure \ref{fig-cp_prebisimGame} when challenger chooses to play from the $p$ side, when it is checked if $p \precsim_{cp} q$.}
\label{tab-prebisimZ_A}
\end{table}

\begin{table}[t]\footnotesize
\begin{tabular}{|c|m{5cm}|m{8cm}|} \hline
\multicolumn{3}{|c|}{Challenger Selects $Z_B$} \\ \hline
& Challenger moves & Defender moves (on $Z_A$) \\ \hline
Move 1 & $(L_0, y = 0) \to{a} (L_1, y = 0)$ & $(l_0, x = 0) \to{a} (l_1, x = 0)$ \\ \hline
\multirow{2}{*}{Move 2} & i) $(L_0, y = 0) \to{11} (L_1, y = 11)$ & $(l_0, x = 0) \to{11} (l_0, x = 11)$ (Note that $(l_0, x = 4)$ is not prebisimilar to $(L_0, y = 11)$.) \\ \cline{2-3}
& ii) $(L_0, y = 11 + \delta) \to{4 - \delta} (L_1, y = 15)$ (From ($L_0, y = 11$), the challenger can make a move to ($L_0, y = 11 + \delta$)) & $(l_0, y = 11 + \delta) \to{1 - \delta} (l_0, x = 12)$. (From $(l_0, x = 11)$, the defender also makes a $\delta$ move to reach $(l_0, x = 11 + \delta)$.)  \\ \hline
\multirow{3}{*}{Move 3} & i) $(L_0, y = 0) \to{11 + \delta} (L_0, y = 11 + \delta)$ & $(l_0, x = 0) \to{11 + \delta} (l_0, x = 11 + \delta)$  \\ \cline{2-3}
& ii) $(L_0, y = 11 + \delta) \to{4} (L_0, y = 15 + \delta)$ & $(l_0, x = 11 + \delta) \to{1} (l_0, x = 12 + \delta)$ \\ \cline{2-3}
& iii) $(L_0, y = 15 + \delta) \to{5} (L_0, y = 20 + \delta)$ & $(l_0, x = 12 + \delta) \to{4 - \delta} (l_0, x = 16)$ \\ \hline
\end{tabular}
\caption{Moves of the cp-prebisimulation game corresponding to automata shown in Figure \ref{fig-cp_prebisimGame} when challenger chooses to play from the $q$ side, \delete{where}when it is checked if $p \precsim_{cp} q$.}\label{tab-prebisimZ_B}
\end{table}
Similar to corner point bisimulation, in cp-prebisimulation too, only finitely many pairs of states are considered for checking the relation.
\begin{theorem}
Corner point prebisimilarity between two timed automata states is decidable.
\end{theorem}
\begin{theorem} \label{thm-prebisim}
Two timed automata states are timed performance prebisimilar if and only if they are corner point prebisimilar.
\end{theorem}

\subsection{Deciding Time Abstracted Bisimulation}
Time abstracted bisimulation between two timed automata has been shown to be decidable \cite{LAKJ1}\cite{LY1} using the region graph construction \cite{AD1}. Two timed automata are timed abstracted bisimilar if and only if their region graphs are strongly bisimilar. We use the zone graph constructed in algorithm \ref{algo-zonegraph} instead of the region graph. 
The size of the zone graph is independent of the constants with which the clocks are compared in the timed automaton guards. Let $Z_{(A, p)}$ \delete{be}denote the zone graph for \delete{timed}state $p$ of timed automaton $A$. If there are two valuations $(l, v)$ and $(l, v')$ such that they belong to the same node, then by construction of $Z_{(A, p)}$, $(l, v)$ and $(l, v')$ are time abstracted bisimilar. Thus in the zone graph, it is the case that a state $(l, v)$ in $T(A)$ is time abstracted bisimilar to the zone $z$ to which $(l, v)$ belongs in the zone graph $Z_{(A, p)}$. The same holds for a timed state $(l_2, v_2)$ in $T(B)$. Thus checking whether two states $(l_1, v_1)$ and $(l_2, v_2)$ of two timed automata $A$ and $B$ are time abstracted bisimilar involves checking whether their corresponding nodes in the two zone graphs are strongly bisimilar.
The following theorems show how time abstracted delay bisimulation and time abstracted 
observational bisimulation \cite{TY1} too can be decided along with strong time abstracted bisimulation using our zone graph.
\begin{theorem}\label{delaytab}
Let $\mathcal{R} \subseteq S_1 \times S_2$ be a symmetric relation. Two nodes ($s_1, s_2$) $\in \mathcal{R}$ if and only if
$\forall a \in Act, \forall s_1' [s_1 \stackrel{a}{\rightarrow}s_1' \Rightarrow \exists s_2' \:.\:  s_2 \stackrel{\beta}{\rightarrow}s_2'$ and $(s_1', s_2') \in \mathcal{R}]$ and
$\forall s_1', [s_1 \stackrel{\varepsilon}{\rightarrow}s_1' \Rightarrow \exists s_2' \:.\:  s_2 \stackrel{\varepsilon}{\rightarrow}s_2'$ and $(s_1', s_2') \in \mathcal{R}]$ \\
Two states $p$ and $q$ are time abstracted bisimilar iff $Z_{(A, p)}\: \mathcal{R}\: Z_{(B, q)}$ and $\beta$ is the action $a$, \delete{they}$p$ and $q$ are time abstracted delay bisimilar iff $Z_{(A, p)}\: \mathcal{R}\: Z_{(B, q)}$  and $\beta$ is the sequence of actions $\varepsilon.a$ whereas $p$ and $q$ are time abstracted observational bisimilar iff $Z_{(A, p)}\: \mathcal{R}\: Z_{(B, q)}$  and $\beta$ is the sequence $\varepsilon.a.\varepsilon$.
\end{theorem}
\subsection{Complexity} In our work, we decide the timed and the time abstracted relations using a zone graph approach. For a given location, the zones in the zone graph are disjoint. In the worst case, the size of the zone graph is limited by the size of the region graph and it is thus exponential in the number of clocks of the timed automaton. However, in \delete{many}most cases, the size of the zone graph is much smaller than the size of the region graph. Existing approaches for checking timed bisimulation involve a product construction on the region graphs or zone graphs
which characterizes the common behaviour of the two timed automata. The product, along with the individual region graphs or zone graphs is stored in order to check for timed simulation relation or timed bisimulation.  

In our case, we do not use a product construction on the zones and use the individual zone graphs of the two timed automata directly for deciding the relations. Thus our method is more space efficient than the approaches that store the product of the region graphs or zone graphs for checking timed bisimulation. Deciding timed bisimulation and timed simulation is known to be EXPTIME-complete \cite{LH1}. Thus our algorithm is not asymptotically better than existing approaches, however we expect that 
we will obtain significant performance gains than existing approaches since our method eliminates the product construction which is an expensive operation. For time abstracted relations, the complexity is similar to the method used in \cite{TY1} which too uses strong bisimulation on zone graph.

\section{Game Characterization}\label{sec-game}
Bisimulation games were defined in \cite{CS1} for discrete processes. In \cite{CD1}, Game characteizations have been given for relations in the van Glabbeek spectrum \cite{VG1}. We present here game characterizations for timed relations that is similar to bisimulation games and define the game semantics using our zone graph. As in the bisimulation game, the game is played in rounds on two graphs. The game may be played between the nodes of the zone graph (game for time abstracted relations) or between the timed states appearing in some node of the zone graph (game for timed relations). In each round, the challenger chooses a graph and the defender tries to make a corresponding move on the other graph where the correspondence of the moves is defined in subsection \ref{subsection-template} in terms of the tuple $\alpha$. If the defender can always make a move in response to the challenger's move, then it has a winning strategy implying that the two states are related through the relation that corresponds to the game. Otherwise it loses which implies that the two states are not related in which case the challenger is said to have a winning strategy. \delete{If the challenger changes the graph between two consecutive rounds, it is known as an \emph{alternation}.} An \emph{alternation} occurs if the challenger changes the graph between two consecutive rounds.  Alternations are not allowed in simulation preorder and simulation equivalence games. A game always \emph{terminates} due to the finiteness of the zone graph and due to the fact that the moves of the game are not repeated from a pair of points that have been visited earlier. In the games described in this section, a \emph{move} is a visible action or a delay action or a sequence of actions where each action belongs to the set $Act \cup \{\varepsilon\}$.
\newcommand{\set}[1]{\{{#1}\}}
\subsection{Game Template} \label{subsection-template}
\vspace{-10pt}
A timed game proposed in this work can be described as $n-\Gamma_{k}^{\alpha, \beta}$. The timed performance prebisimulation game (cp-prebisimulation game) consists of two parts where the second subgame is played if the defender loses in the first subgame. Either of the subgames can be played first and hence the two subgames are connected by a $\vee$. Each game is characterized by the following parameters:
\begin{itemize}
\item[-] $n$ : number of alternations. If not mentioned, then there is no restriction on the number of alternations.
\item[-] $k\in\set{\mathbb{N}\cup \infty}$ : number of rounds; $n\le k-1 \text{ when } k \neq \infty$.
\item[-] $\alpha$ : a tuple $\langle \alpha_1, \alpha_2 \rangle$. $\alpha_1$ denotes the move chosen by the challenger. Depending on the game for the timed relation, either $\alpha_1 \in Lep$ or $\alpha_1 \in Act \cup \realpos$ whereas $\alpha_2$ denotes the move chosen by the defender and may be the same as $\alpha_1$ or may be a sequence of the form $\varepsilon . \alpha_1$ or $\varepsilon . \alpha_1 . \varepsilon$, as in the case of time abstracted relations. For example, for the pair $\langle a / \varepsilon, \varepsilon . a / \varepsilon \rangle$ where $a \in Act$, the challenger makes a move $a$ whereas the defender's move consists of $\varepsilon$ followed by an $a$. In the case of timed bisimulation game (or cp-bisimulation game), $\alpha$ is assigned $\langle a/d, a/d \rangle$, which denotes that a visible action by the challenger has to be matched by the defender and a delay action $d$ by the challenger has to be matched with an exact delay $d$ move by the defender. In a timed performance prebisimulation game or (cp-prebisimulation game), $\alpha$ is assigned $\langle a/d_1, a/d_2 \rangle$ denoting that the delays performed by the challenger and the defender need not be the same.
\item[-] $\beta$ : This is an extra condition which is used in the cp-prebisimulation game. When the game is played between the zone graphs $Z_{A}$ and $Z_{B}$ corresponding to the two timed automata $A$ and $B$, $Z_{A}, \le$ denotes that the delay moves made in $Z_{A}$ are no more than the delays made in $Z_{B}$. $\beta$ if  not specified denotes that there is no extra condition.
\end{itemize}

\xyoption{curve}
\newcommand{\HierarchyFigureWithSpecturm}{
\begin{wrapfigure}{rt}{0.5\textwidth}
\vspace{-10pt}
\scriptsize	
\begin{tabular}{l|l}
\subfigure[]{
$\xymatrix @R=15pt @C=0pt @M=0pt{
& \txt{Timed Bisimulation} \ar@{->}[d] \ar[dl]\\
\txt{Timed Simulation\\ Equivalence} \ar[dd]
& \txt{Timed Performance\\ Prebisimulation} \ar[d] \\
& \txt{Time Abstracted\\ Bisimulation} \ar[d] \ar[dl]\\
\txt{Time Abstracted\\ Simulation Equivalence} \ar[d] & \txt{Time Abstracted\\ Delay\\ Bisimulation} \ar[d] \ar[dl] \\
\txt{Time Abstracted\\ Delay Simulation\\ Equivalence} \ar[d] &  \txt{Time Abstracted\\ Observational\\ Bisimulation} \ar[dl] \\
\txt{Time Abstracted\\ Observational \\Simulation Equivalence} & \\ \\
}$
\label{fig-timedSpectrum} 
}&
\subfigure[]{
$\xymatrix @R=24pt @C=10pt @M=0pt{
& {\TGame{}{\infty}{\langle a/d,a/d \rangle}} \ar[d] \ar@/_/[dl]&\\
{\TGame{0}{\infty}{\langle a/d,a/d \rangle}} \ar[dd] & \txt{$\TGame{}{\infty}{\langle a/d,a/d \rangle, (Z_{A}, \le)}$ \\ $\lor \: {\TGame{}{\infty}{\langle a/d,a/d \rangle, (Z_{A}, \le)}}$} \ar[d] \\
& {\TGame{}{\infty}{\langle a/ \varepsilon, a / \varepsilon\rangle}} \ar[d] \ar[dl]&\\
{\TGame{0}{\infty}{\langle  a/ \varepsilon, a / \varepsilon \rangle}} \ar[d] & {\TGame{}{\infty}{\langle a/ \varepsilon,\varepsilon . a / \varepsilon \rangle}} \ar[d] \ar[dl]& \\
{\TGame{0}{\infty}{\langle a/ \varepsilon,\varepsilon . a / \varepsilon \rangle}} \ar[d] &  {\TGame{}{\infty}{\langle a / \varepsilon,\varepsilon . a . \varepsilon / \varepsilon\rangle}}  \ar[dl]& \\
{\TGame{0}{\infty}{\langle a / \varepsilon,\varepsilon . a . \varepsilon / \varepsilon\rangle}}  & 
}$
\label{fig-gamehierarchy}
}
\end{tabular}
\vspace{-10pt}
\caption{\subref{fig-timedSpectrum} presents the spectrum of timed relations and \subref{fig-gamehierarchy} shows timed games corresponding to these relations}
\label{fig:subfigureExample}
\end{wrapfigure}
\vspace{-10pt}
}
\subsection{Hierarchy of Timed Games} \label{sec-gamehier}
A hierarchy among the timed relations discussed in this paper is captured in Figure \ref{fig-infgamehierarchy}(a). We show here several lemmas which capture this hierarchy through the game semantics. These lemmas also help us build an infinite game hierarchy which also suggests defining several new timed relations that do not exist in the literature. The arrow from a game $\Gamma_1$ to a game $\Gamma_2$ denotes that if the defender has a winning strategy for $\Gamma_1$, then it also has a winning strategy for $\Gamma_2$. Besides in each of the following lemmas, for each pair of games, if $\Gamma_1 \longrightarrow \Gamma_2$, then $\Gamma_2 \not \longrightarrow \Gamma_1$. Figure \ref{fig-infgamehierarchy}(b) shows the games corresponding to the relations shown in Figure \ref{fig-infgamehierarchy}(a). The game hierarchy reflects the hierarchy of the timed relations.

\begin{lemma}
$\Gamma_{\infty}^{\alpha, \beta} \longrightarrow n\!-\!\Gamma_{\infty}^{\alpha, \beta} \longrightarrow (n\!-\!1)\!-\!\Gamma_{\infty}^{\alpha, \beta}$, for all $n > 0$ \\
$\Gamma_k^{\alpha, \beta} \longrightarrow n\!-\!\Gamma_k^{\alpha, \beta} \longrightarrow (n\!-\!1)\!-\!\Gamma_k^{\alpha, \beta}$, for all $k > 0$, $n < k$
\end{lemma}
Other parameters remaining the same, if the defender has a winning strategy when the challenger is allowed more alternations, then the defender also wins the game where the challenger is allowed only a smaller number of alternations.
\begin{lemma}
$\Gamma_{\infty}^{\alpha, \beta} \longrightarrow \Gamma_{k}^{\alpha, \beta} \longrightarrow \Gamma_{k-1}^{\alpha, \beta}$, for all $k > 0$ \\
$n\!-\!\Gamma_{\infty}^{\alpha, \beta} \longrightarrow n\!-\!\Gamma_{k}^{\alpha, \beta} \longrightarrow n\!-\!\Gamma_{k-1}^{\alpha, \beta}$, for all $k > 0$, $n < k$
\end{lemma}
Other parameters remaining the same, if the defender wins the game with more number of rounds, then it also wins the game which has a smaller number of rounds in the game.
\begin{lemma}\label{lem-third}
\delete{$n-\Gamma_{k}^{\alpha, =} \longrightarrow n-\Gamma_{k}^{\alpha, \lfloor = \rfloor} $\\}
$n-\Gamma_{k}^{\langle a/d, a/d \rangle} \longrightarrow n-\Gamma_{k}^{\langle a/d_1, a/d_2 \rangle, (Z_{A}, \le)} $.\\
\delete{$n-\Gamma_{k}^{\alpha, =} \longrightarrow n-\Gamma_{k}^{\alpha, (G_2, \le)} $\\}
$n-\Gamma_{k}^{\langle a/d, a/d \rangle} \longrightarrow n-\Gamma_{k}^{\langle a/d_1, a/d_2 \rangle, (Z_{A}, \le)} \: \vee \: n-\Gamma_{k}^{\langle a/d_1, a/d_2 \rangle, (Z_{B}, \le)}$
\delete{$n-\Gamma_{k}^{\alpha} \longrightarrow n-\Gamma_{k}^{\alpha} $}
\end{lemma}
The first half of the above lemma states that all the parameters remaining the same, if the defender can always reply with an exact delay, then the defender can reply with a delay $d_2$ in $Z_{A}$ such that $d_2 \le d_1$ and it can reply with a delay $d_2$ in $Z_{B}$ such that $d_1 \le d_2$. This also leads to the fact that all the parameters remaining the same, if the defender wins the cp-bisimulation game, then it also wins the cp-prebisimulation game.
\begin{lemma} \label{lem-epsilon}
$n-\Gamma_{k}^{\langle a/d, a/d \rangle} \longrightarrow n-\Gamma_{k}^{\langle a / \varepsilon, a / \varepsilon \rangle} \longrightarrow n-\Gamma_{k}^{\langle a  / \varepsilon, \varepsilon . a  / \varepsilon\rangle} \longrightarrow n-\Gamma_{k}^{\langle a / \varepsilon, \varepsilon . a . \varepsilon  / \varepsilon\rangle}$
\end{lemma}
If the defender can match a delay action exactly as in the corner point bisimulation, then it can match an epsilon move of the challenger. Also if the defender can reply to a visible action of the challenger, then it can reply with an $\varepsilon.a$ or an $\varepsilon.a.\varepsilon$ move since $\varepsilon$ \delete{may}represents delay including \delete{a}zero delay.

\subsection{Infinite Game Hierarchy}
\begin{figure}[h]
\centering
\includegraphics[width=.7\textwidth]{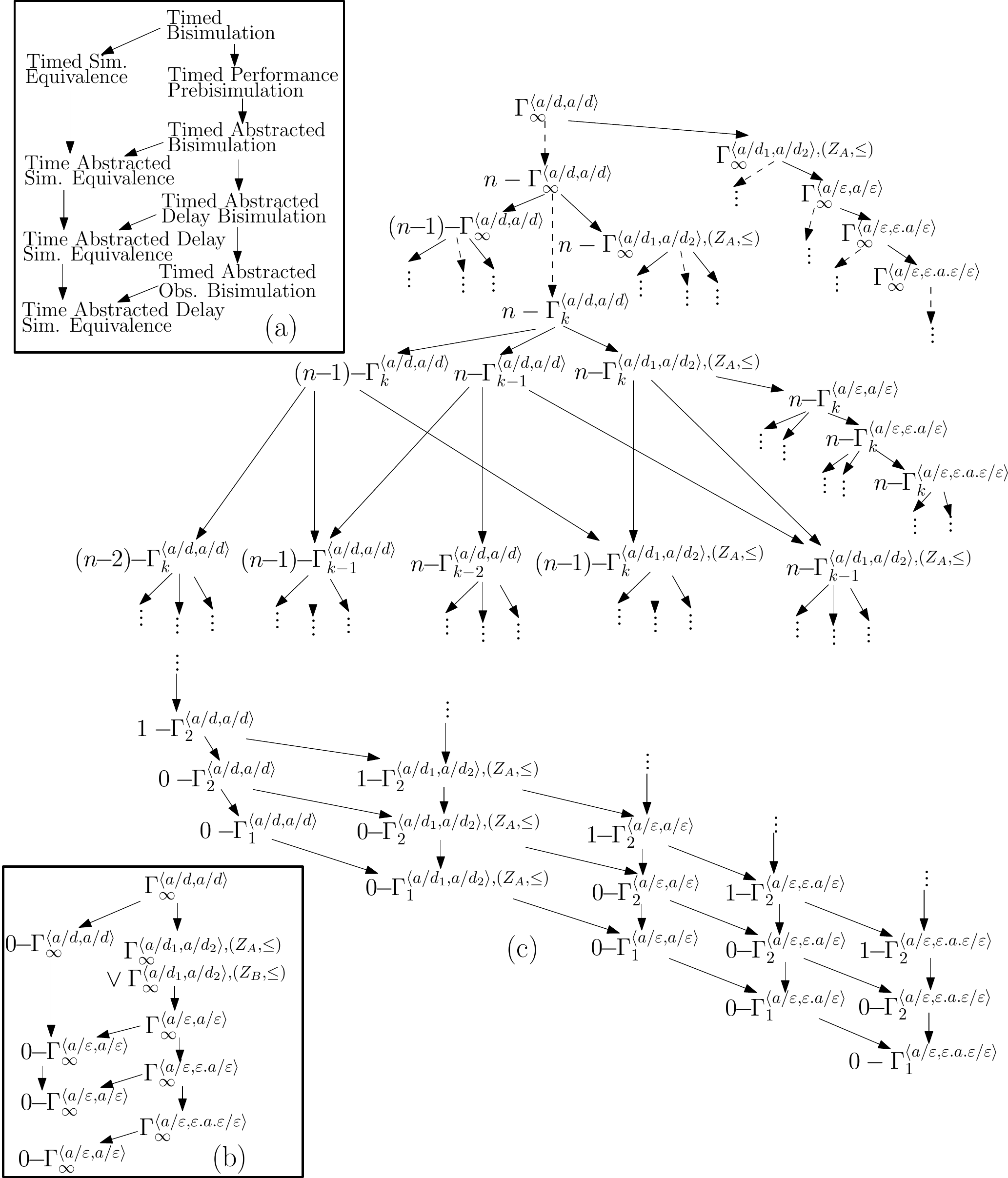}
\vspace{-5pt}
\caption{\label{fig-infgamehierarchy} Relations over timed automata, game characterization and the infinite hierarchy of timed games}
\vspace{-10pt}
\end{figure}

On assigning different values to the parameters $n$, $k$, $G$, $\alpha$ and $\beta$ in the game template and using the lemmas given in subsection \ref{sec-gamehier}, we can generate an infinite game hierarchy which is shown in Figure \ref{fig-infgamehierarchy}(c). The dashed lines in the figure denote that if the defender has a winning strategy for a game with infinitely many rounds or alternations, then it also wins a game with a finite number of rounds or alternations. Figure \ref{fig-infgamehierarchy}(b) shows the hierarchy of the games that correspond to the timed relations in Figure \ref{fig-infgamehierarchy}(a).
The diagram in Figure \ref{fig-infgamehierarchy}(b) is only a small part of the entire hierarchy of timed games and this leaves us with the scope of studying several timed relations that are not present in the existing literature.
\vspace{-10pt}
\section{Conclusion}\label{sec-conc}
\vspace{-5pt}
In this paper, we present a unified zone based approach to decide various timed relations between two timed automata states. In our method, we do not need the product construction on regions or zones for deciding these relations as done in \cite{KC1} or \cite{WL1}. We also provide a game semantics for deciding these timed relations and show that the hierarchy among the games reflects the hierarchy among the relations. The advantage of a game-theoretic formulation is that it allows fairly general relationships between the parameters on $\Gamma$ to define the hierarchy. The fine-tuning and variations of these parameters allow formulations of many more equivalences and preorders than the ones present in the literature related to behavioural equivalences involving real time which otherwise may not be easily captured through operational definitions and reasoning. 
Unlike existing approaches which check if two timed automata states are related through some relation, our game approach also allows generating a distinguishing formula that guides us to find a path in one of the zone graphs which was responsible for the relation not holding good between the corresponding states. Identifying this path helps us to refine appropriately an implementation that should conform to a given specification through the relation.
As further work, we plan to extend the game semantics to relations over probabilistic extensions to timed automata \cite{KNSS1}. We would also like to investigate the applicability of our zone graph construction for deciding these relations.
\vspace{-20pt}
\bibliographystyle{eptcs}
\bibliography{express}

\end{document}